# Water jet rebounds on hydrophobic surfaces : a first step to jet micro-fluidics.

F. Celestini *, R. Kofman, X. Noblin and M. Pellegrin.

*Laboratoire de Physique de la Matère Condensée, CNRS UMR 6622, Université de Nice Sophia Antipolis, Parc Valrose, 06108 Nice, France.*

**When a water jet impinges upon a solid surface it produces a so called hydraulic jump that everyone can observe in the sink of its kitchen. It is characterized by a thin liquid sheet bounded by a circular rise of the surface due to capillary and gravitational forces. In this phenomenon, the impact induces a geometrical transition, from the cylindrical one of the jet to the bi-dimensional one of the film. A true jet rebound on a solid surface, for which the cylindrical geometry is preserved, has never been yet observed. Here we experimentally demonstrate that a water jet can impact a solid surface without being destabilized. Depending on the incident angle of the impinging jet, its velocity and the degree of hydrophobicity of the substrate, the jet can i) bounce on the surface with a fixed reflected angle, ii) land on it and give rise to a supported jet or iii) be destabilized, emitting drops. Capillary forces are predominant at the sub-millimetric jet scale considered in this work, along with the hydrophobic nature of the substrate. The results presented in this letter raise the fundamental problem of knowing why such capillary hydraulic jump gives rise to this unexpected jet rebound phenomenon. This study furthermore offers new and promising possibilities to handle little quantity of water through "jet micro-fluidics"**

The specular law of reflection (Snell-Descartes) is certainly one of the most famous physical laws. It simply states that a light beam is reflected by a surface with an angle $\theta_r$ equal to the incident one $\theta_i$. Such a general and deterministic law does not exist to describe the reflection of matter on a surface. For example, an isolated atom obeys to the stochastic Lambert's reflection law stating that the probability for an atom to be reflected with an angle $\theta_r$ is proportional to $\cos \theta_r$. In spite of the long-standing interest and activity in the physics of liquid jets (1), a true jet reflection preserving the cylindrical geometry of the incident stream of matter has never been observed and consequently a reflection law was never yet proposed. Even if recent studies have demonstrated that the hydraulic jump (HJ) can deviate from its classical circular shape in the case of normal incidence (2, 3) or for inclined impinging jets (4), in both cases the fluid stays on the surface with a planar geometry. Until now, there are only two situations for which a jet preserves its geometry after an impact: the first one is known as the Kaye effect (5) for a viscoelastic fluid and the second has been recently discovered (6) for a Newtonian liquid. In both cases, the jets impact on the same liquid.

In the present work, we report the experimental observation of a water jet bouncing on a solid surface. The impact induces a capillary hydraulic jump (CHJ) somehow similar to the HJ in the sense that the symmetry is broken (jet towards liquid sheet) and that a thin liquid sheet is created. The main difference comes from the fact that capillarity is the driving force that allows the jet to take off from the surface while for a HJ gravitational and capillary forces induce the abrupt rise of the water height. Our observations are indeed made for sub-millimetric jets impinging hydrophobic surfaces such that capillary forces are predominant over gravity. Jets are created by pushing water into nozzles of radius R. We perform experiments with radii R=84, 197 and 270 μm. The velocity of the incident jets is controlled and, depending on the radius and the applied pressure, varies between 1 and 5 ms$^{-1}$ corresponding to Reynolds numbers in between 100 and 3000. For the small radii and the free fall distances (typically a few mm) considered, the gravity can be neglected and the jet velocity at the end of the nozzle can be considered to be the same as the one at the impact on the solid surface. We also control the angle between the incoming jet and the substrate. This incident angle (Fig. 2a) is denoted $\theta_i$ and, when observed, the reflected one is denoted $\theta_r$. The incident angle can be varied from 0, corresponding to a normal incidence, up to 85 degrees. We present in this paper the results obtained for impacts on two different hydrophobic (H) and super-hydrophobic (SH) solid surfaces. The hydrophobic surface is made of a Teflon coating (Teflon AF 1600, DuPont) obtained by evaporation on a silicon wafer. The contact angle is of 110°. The superhydrophobic surface is obtained by coating a thin Teflon film (sub-micronic) on a micro structured surface. Using a photolithography technique we have made cylindrical micropillars arrays in SU-8 photoresist (Microchem) on a square lattice (40 microns pitch). We obtain a contact angle of 155 °. In this article, we first present general observations of the jet rebound and we describe the jet oscillations after impact. Then, we analyse the rebound characteristics in terms of restitution coefficients and we propose a framework for a theoretical model that explains the bouncing phenomenon. Finally, we illustrate how fine jet control can be achieved and we present experimental perspectives.

An example of a jet rebound is given in Fig. 1a for a jet of radius R=0.197 mm and an incidence $\theta_i = 68°$ on the SH surface. We can see the CHJ and the reflected jet displaying stable oscillations. These oscillations are put in evidence by the shadow of the reflected jet on the substrate as well as by the refracted light. This situation of a stable bouncing jet is not generic. Depending on the velocity, on the incident angle and on the nature of the solid surface, the situation can be rather different. This is what is shown in Fig. 1b where we report the phase diagram obtained for the H and SH surfaces. For small velocities, the jet can literally land on the surface giving rise to a what we named "supported" jet, in analogy with the terminology used for drops deposited on a substrate. This situation is interesting insofar as it shows the possibility to guide the jet on a solid surface since it follows a straight direction at least on a small distance. At larger distances from the impact point, the landing jet gives rise to a meandering (7). The picture (Fig. 1b) illustrating the landing jet has been taken for a SH surface. For the H surface, the landing jet can also display strong oscillations similar to those of the reflected jet. This will be discussed below and illustrated in Fig. 4a. For the largest velocities, the jet is destabilized and emits droplets. We did not identify precisely the frontier between landing and destabilized jets but we focused on the region where the reflected jet is stable. One can clearly see that this region is larger for the SH than for the H substrates considered. The degree of hydrophobicity of the solid surface is a crucial parameter for the bouncing jet. It can be measured through the area of the stable bouncing jet region in units of velocity times degrees. This measure is directly connected to both wetting angle and hysteresis of the surface. Thus, it globally characterises the surface through its ability to create stable rebounds.

We first focus on the oscillations appearing in the reflected jet. For the three nozzles and different incident velocities V, we have measured the wavelength $\lambda$ (Fig. 2a) of the oscillations. As discussed above, they are stable so that we can extract a characteristic oscillation time $T_\lambda$ of a cross section of the liquid jet. This time can simply be measured through the relation: $T_\lambda = \lambda/V'$, V' being the reflected velocity evaluated by measuring the radius R' of the reflected jet and using mass conservation: $V'=R^2V/R'^2$. The mean value of the reflected jet is obtained averaging between maximum and minimum values caused by the oscillations. A top view of the CHJ is given in Fig. 2b, its shape is rather similar to the ones observed for inclined jets (4). When the jet takes off, it escapes from a bi-dimensional geometry to return to a cylindrical one minimizing its surface energy. Since the Reynolds number is largely greater than unity, the jet displays non-axisymmetric oscillations in a similar manner than a jet discharging from an elliptical nozzle does. Rayleigh (8) and Bohr (9) described this phenomenon and proposed an analytical expression for the oscillation time: $T_\lambda = 2\pi (\rho R'^3/6\gamma)^{1/2}$ where R' is the radius of the reflected jet, $\rho$ the density and $\gamma$ the surface tension. As shown in Fig. 2c, this expression fits very well our experimental data without any free parameter. The oscillations decrease in amplitude as the distance from the impact increases and the jet therefore turns back to its initial cylindrical shape. Depending on the velocity, the incident angle and the hydrophobicity of the solid substrate, the oscillations can be more or less important. The jet rebound can therefore be viewed as an efficient way to produce jets similar to the ones discharging from elliptical nozzles with tuneable asymmetries. As for the decrease in amplitude, the asymmetry is controlled by the properties of the incident jet. This is illustrated in Fig. 2a and 2b where the jet is in contact with the substrate over a distance *d* that is the longitudinal extension of the CHJ.

As done just above for the oscillation of the bouncing jet, we can extract the quantity $T_d = d / V \sin\theta_i$ giving the contact time of the jet on the substrate. The contact time $T_d$ is approximately evaluated assuming that the incident jet parallel velocity $V\sin\theta_i$ is not modified during the contact. We plot in Fig 3a. the values of this contact time normalized by $T_\lambda$ as a function of the perpendicular Weber number $W_e^\perp = \rho R V_\perp^2 / \gamma$, measuring the relative importance of kinetic versus capillary effects. The normalized contact time is represented for the different jet radii, velocities and incident angles considered in this study. As can be seen in Fig. 3a, the overall data merge on a same curve. For small Weber numbers, the contact time is almost the same as the oscillation time. When kinetic effects are larger, we can observe that $T_d$ increases and that, for the higher Weber numbers, it can be twice greater than $T_\lambda$. It is important to note that such a collapse of all the data in a single curve can solely be obtained using the perpendicular Weber defined above. The perpendicular component of the velocity changes is direction of 180 degrees. It is therefore the main parameter that controls the bouncing of the jet.

The results presented above lead us to propose a simple phenomenological interpretation based on an analogy where the jet is considered as a "chain of springs" (Fig. 3b). Each spring models a small piece of water and its stiffness mimic the effect of the surface tension. The equilibrium position of the spring corresponds to a jet with a cylindrical geometry as it is before the impact. This approach is similar to the one that has been successfully applied to the rebounds of water drops (10). It is nevertheless much more restrictive in the present case since we neglect all interactions between springs. In other words, we do not consider the effect of deformations parallel to the velocity of the jet, but solely perpendicular deformations. In spite of its simplicity, the analogy gives a good physical understanding of the oscillations of the reflected jet. The impact induces a compression of the springs that take off from the substrate when their perpendicular acceleration is maximal. The inertia is at the origin of the observed oscillations after the take-off. According to the work on drop rebounds, the impact time should be equal to the eigen period of the spring. As can be seen in Fig. 3a, we recover a similar behaviour for the jet rebound where, for small incident perpendicular velocities, the contact time tends to the oscillation time of the jet. The increase of $T_d$ can be interpreted as a non-linear effect that shifts the resonant frequency of the jet submitted to large deformations. As we can observe in Fig. 2b, a thin liquid sheet separates two liquid rims in the central zone of the CHJ. To model such a situation, one should consider the full hydrodynamics equations but this would go beyond the purpose of this paper. Such a theoretical study has been proposed for a similar system where two jets collide and create a so-called fishbone structure (11).

As for the bouncing of a ball or a drop (10), we have measured both perpendicular $\alpha_\perp = \dfrac{V'_\perp}{V_\perp}$ and parallel $\alpha_{/\!/} = \dfrac{V'_{/\!/}}{V_{/\!/}}$ restitution coefficients, the subscripts referring to the perpendicular or parallel components respectively and the superscript to the reflected (') or incident velocities. These coefficients quantify the amount of energy restituted in the two directions. In Fig. 3c. we plot both restitution coefficients as a function of the perpendicular Weber number. The behaviours of the parallel and perpendicular components are qualitatively different. For small velocities, one can see that the restitution coefficient is greater for the parallel component ($\alpha_{//} \approx 1$) than for the perpendicular one and that almost all the energy is lost in the perpendicular direction. For higher velocities, both components are decreasing and their ratio, $r = \alpha_{//} / \alpha_{\perp}$, tends to a constant value. The dashed line corresponds to a regime in which $\alpha_{\perp}$ scales as $1/\sqrt{W_e^{\perp}}$ meaning that the velocity of the take off is independent of the incident velocity as it is the case for drop rebounds at high Weber numbers. As for every rebound events, the main question is to know the reflected angle $\theta_r$ and its dependence on the incident one $\theta_i$. We plot in Fig 3d the values of $\theta_r$ as a function of $\theta_i$. The circles correspond to the angles obtained for a stable bouncing jet on the SH substrate while dots represent the overall data obtained. In all cases, the reflected angle is greater than the incident one. The reflected angle can be expressed as a function of the restitution coefficients defined above: $\tan\theta_r = r \tan\theta_i$. Using the value found for r at high incident velocities, the agreement is satisfactory at least for reflected angles of stable jets.

In order to obtain theoretical expressions for the restitution coefficients and to explain the reflection angles observed, a simple way is to write momentum conservation along both directions (longitudinal: $e_x$ and vertical: $e_y$). We consider a control surface including the liquid sheet and both incident and reflected jets, bordered by the fluid-air interface and the fluid-substrate surface ($\Sigma$). By projecting on the $e_x$ direction the momentum conservation, we obtain:

$$-\pi \rho V^2 R^2 \sin\theta_i + \pi \rho V'^2 R'^2 \sin\theta_r = -F_{//} \text{ with } F_{//} = \int_{\Sigma} \eta \frac{\partial V}{\partial y} dS$$

$\eta$ being the fluid viscosity. Using mass conservation, the parallel restitution coefficient writes:

$$\alpha_{//} = 1 - \frac{F_{//}}{\pi \rho V^2 R^2 \sin \theta_i}.$$

The horizontal viscous force $F_{//}$ scales as $\frac{\eta S V}{h}$, S being the surface of the liquid sheet and h its characteristic thickness. As we can observe on Fig 3c, $\alpha_{//}$ is found to be close to 1 at low velocities and slightly decreases due to the viscous force. Projecting the momentum conservation on $e_y$ gives the following relation:

$$\pi \rho V'^2 R'^2 \cos \theta_r = -\pi \rho V^2 R^2 \cos \theta_i + \int_\Sigma p dS - F_\perp - \gamma \Pi \sin \theta_e$$

where $\Pi$ is the contact line perimeter. The integral of the pressure forces on the contact surface is a reaction force and is at the origin of bouncing. For a "perfect bouncing event", we have only the substrate reaction force (no surface tension and no dissipation), this term tends to $2\pi \rho V^2 R^2 \cos \theta_i$ and $\theta_i$ tends to $\theta_r$. The third term in the right-hand side corresponds to the vertical viscous dissipations. The last term corresponds to the force due to the surface tension applied all over the perimeter $\Pi$. As we can see, this force tends to spread the jet on the surface and is proportional to the sinus of the equilibrium wetting angle $\theta_e$. We do not intend, in this letter, to derive the exact ratio of the two restitution coefficients. It is influenced by the dependence of the perimeter $\Pi$ and the contact surface $S$ that have not yet been fully examined. Nevertheless, in the limit of small dissipation and at low $W_e$ numbers (so that $\alpha_{//}=1$), the perpendicular restitution coefficient is $\alpha_\perp \approx 1 - \frac{\Pi \sin \theta_e}{W_e \pi R \cos \theta_i}$. In this expression, we quantify the effect of the substrate hydrophobicity on the reflected angle: in a perfect non-wetting situation ($\sin \theta_e = 0$), the reflected angle is equal to the incident one. Finally, we can note that inserting the dependence of $\Pi$ and $S$ should also lead to an expression for the threshold in the incident angle at which the jet rebound appears.

The jet rebound phenomenon presented in this communication opens many new perspectives in the field of manipulation of small liquid quantities. We give three illustrative examples. In the first one (Fig. 4a), the substrate has two different, H and SH, parts. The jet is impinging the H part of the substrate with a velocity and incident angle values chosen so that they lead to the creation of a landing jet. When the latter reaches the SH part of the substrate, it finally takes off and turns to its initial cylindrical geometry. The second example (Fig. 4b) demonstrates that two successive jet rebounds are possible on two SH substrates. In between the two impacts, the jet displays oscillations that are stable in time, as they are for a single rebound. Finally, the previous situation is generalized and we show in Fig. 4c that it is possible to guide a water jet in between two SH substrates somehow as the light is guided in an optical fiber. The two substrates are distant of 1.2mm. It is worth mentioning the difference due to the fact that the jet rebound law is not the same than the specular one standing for light reflection.

The promising results obtained suggest a need for further analysis on two levels. First, on a fundamental point of view, an exact hydrodynamic theory is needed to determine exactly the reflection law as a function of the different parameters, although our model capture the essential features that lead to the jet rebound. Second, as illustrated in the last figure, the presented results should be the starting point for applications in the field of the "jet microfluidics" that has not yet been explored.

* Correspondence should be addressed to Franck Celestini  (franck.celestini@unice.fr)

Acknowledgment: We would like to thank J. Rajchenbach and C. Raufaste for fruitful discussions.

Figure captions:

Figure 1: a) A water jet impinging and bouncing on a super-hydrophobic surface. The reflected jet displays stable non-axisymmetric oscillations similar to the ones of a jet discharging from an elliptical nozzle. These oscillations are brought out by the jet shadow and the light refracted on the substrate. b) Depending on its velocity and on its incident angle, the jet can land, bounce or be destabilized (producing water drops). The phase diagram is given for two substrates with two different equilibrium wetting angles $\theta_e$: a hydrophobic surface made of PTFE ($\theta_e=110°$) and a super-hydrophobic surface made of PTFE coating a structured substrate of micro pillars ($\theta_e=155°$).

Figure 2: a) Side view of the bouncing jet. b) Top view of the capillary hydraulic jump. c) Oscillation periods deduced from the ratio of the wavelength and the velocity of the reflected jet. The full line corresponds to analytical expression without any free parameter.

Figure 3: a) Ratio of the contact ($T_d$) and oscillation ($T_\lambda$) times as a function of the perpendicular Weber number. b) Schematic representation of the analogy between the reflected jet and a chain of springs. c) Parallel (open circles) and perpendicular (filled circles) restitution coefficients as a function of the Weber number associated to the perpendicular component of the incident velocity. Inset: ratio of the restitution coefficients. d) Reflected angle as a function of the incident one. The dotted line corresponds to the specular law of reflection, the full one to the equation $\tan(\theta_r) = r \tan(\theta_i)$ with a value of r equal to 2.1. Circles correspond to stable bouncing jets on the

SH substrate. Dots correspond to the measurements made for all substrates, jet radii and velocities considered in the study. They therefore also include reflected angles of drops when the jet is destabilized.

Figure 4: a) The jet can land on the hydrophobic surface and take off when arriving on the super-hydrophobic substrate. Note that the take-off is very sensitive to the location of the H-SH border. b) A double rebound on two SH substrates. After each rebound, the jet displays stable and decreasing oscillations. c) Multiple jet rebounds (more than five in this case) between two SH substrates separated by a distance of 1.2 mm. The jet is guided between the plates somehow as the light is guided in an optical fiber.

Figure 1.

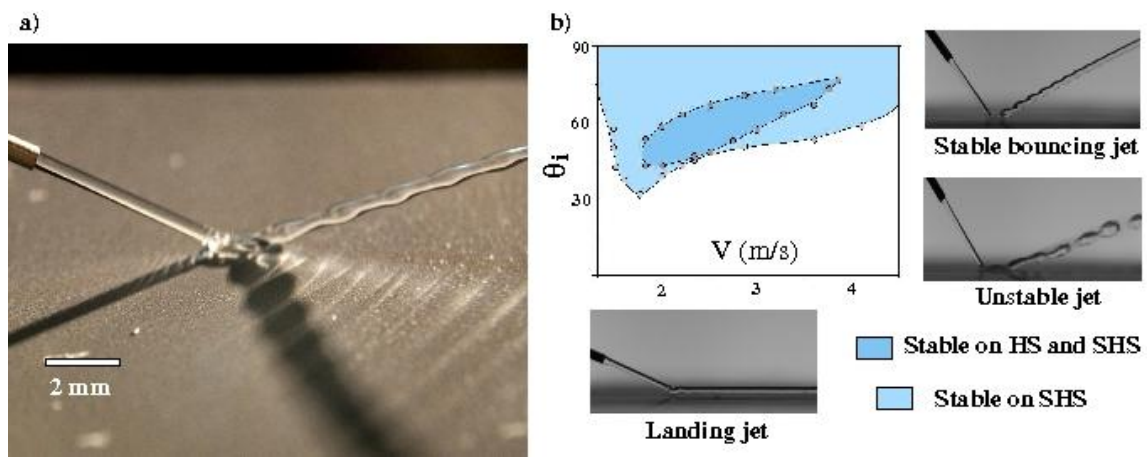

Figure 2.

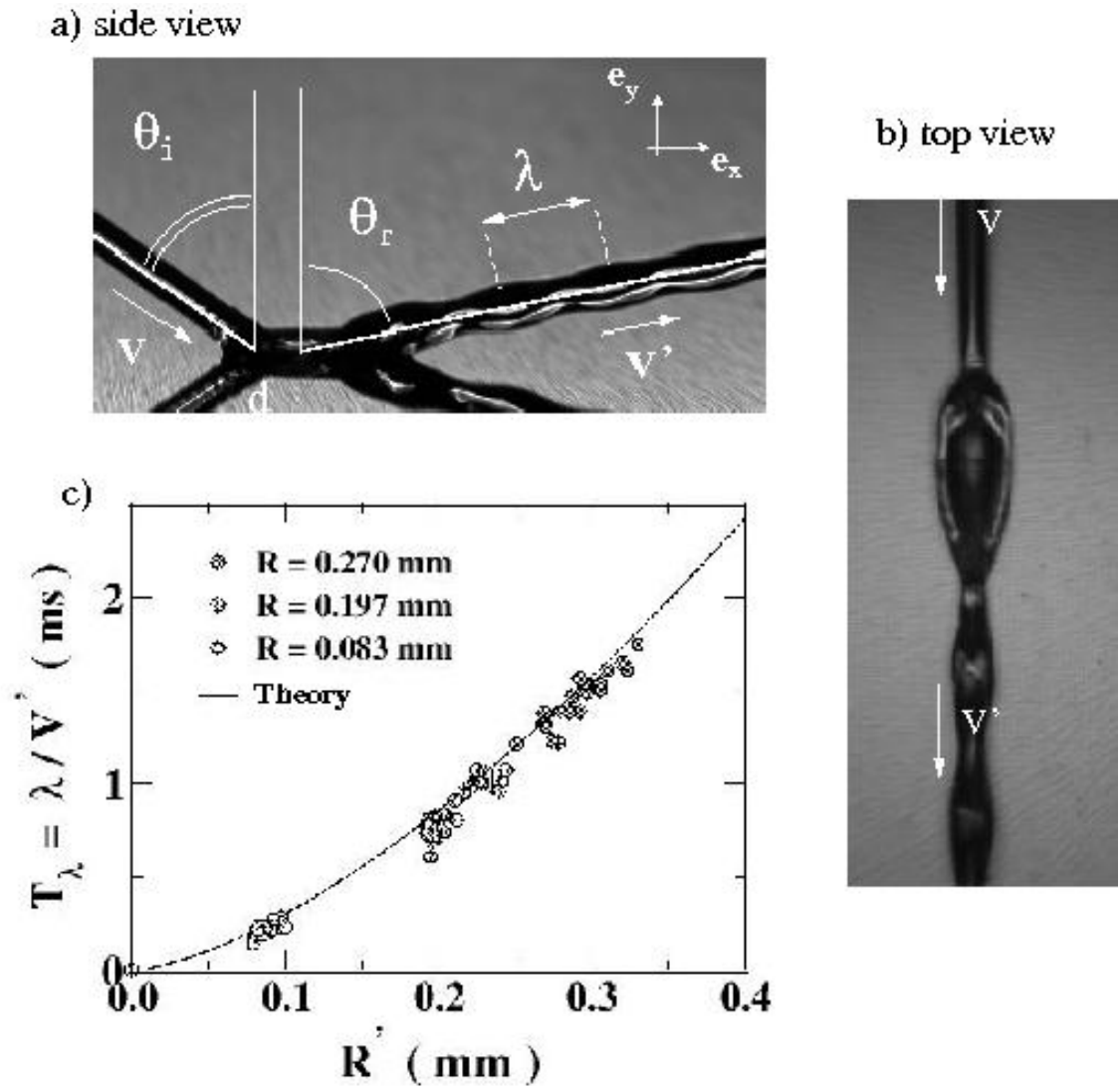

Figure 3.

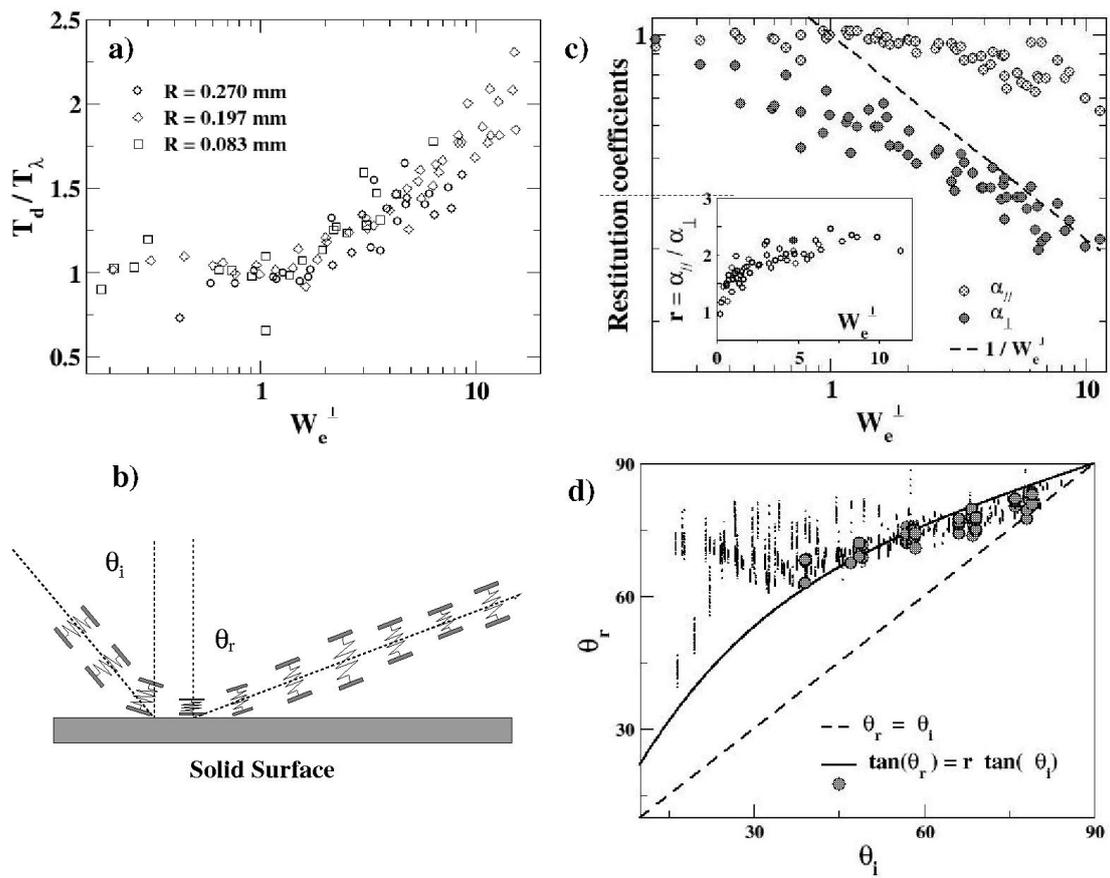

Figure 4.

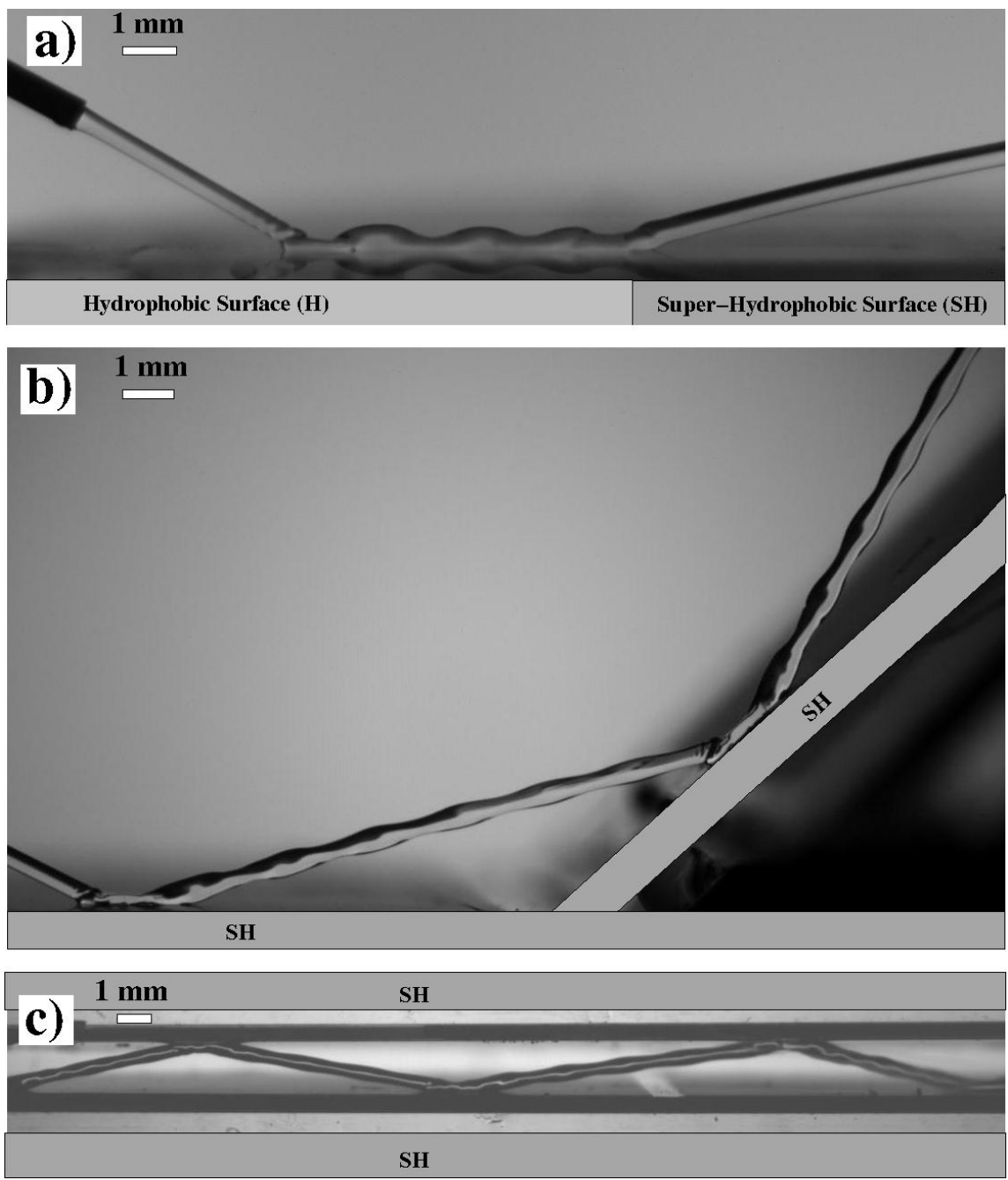